# The proposal of Improved Inexact Isomorphic Graph Algorithm to Detect Design Patterns

Afnan Salem Ba-Brahem, M. Rizwan Jameel Qureshi
Faculty of Computing and Information Technology, King Abdulaziz University, Jeddah, SAUDI ARABIA
e-mail: fonon11_cs_ksa@hotmail.com, anriz@hotmail.com

## Abstract

Design patterns being applied more and more to solve the software engineering difficulties in the object oriented software design procedures. So, the design pattern detection is widely used by software industries. Currently, many solutions presented to detect the design pattern in the system design. In this paper, we will propose a new one which first; we will use the graph implementation to implement both the system design UML diagram and the design pattern UML diagram. Second, we will implement the edges for each one of the both two graphs in a set of 4-tuple elements. Then, we will apply a new inexact graph isomorphic algorithm to detect the design pattern in the system design.

## 1. Introduction

Design patterns are descriptions of the equivalent Classes, which is the appropriate solution to the recurring problems in the design. Once we implement number of patterns when designing a system, the related information to the design pattern will not be longer available. Therefore, it is essential to detect the design pattern instances in the system design to understand the system or maintain it [1-3].

The design patterns regarded as an expert design experiences. The experts' knowledge is represented in the formal 23 patterns of GoF. Also, design pattern are considered as one of the reuse design techniques. Then, the software industry have been reused these experiences widely. As a result of the presence of a reliable method to detect the design patterns from system design, then novices will be success in the software reengineering [4-7].

Many software engineering problems solved by the graph based methods. Also, Design patterns have proved to be effective in solving the problems related to the system design. to detect the design pattern in the system design using the graph based approach we must firstly convert the UML diagram for both: design pattern and the system design into direct graph and then determine whether the pattern graph exist exactly in the system graph or not depend on the classic concept of graph Isomorphic. Sometimes, we do not need to implement the design pattern completely in the system design. In some cases we need to know, how many times a design pattern is completely or partially existed in the system design [1-7].

In this paper, we proposed a new solution to detect the different cases of existence of design pattern in system design. First, we will use the graph implementation to implement both the system design UML diagram and the design pattern UML diagram. Second, we will implement the edges for each one of the both two graphs in a set of 4-tuple elements. Subsequently, we will apply a new inexact graph isomorphic algorithm to detect design pattern in system design.

The rest of this paper is organized as follow: the related work will reviewed in section 2. In section 3, we will present the problem definition and the proposed solution. In section 4, we will validate the proposed solution. Finally, we will present the conclusion from this paper and the future work in section 5.

## 2. Related Work

This section will discuss the previous work in detecting the design patterns. To detect the design patterns in the system design we will decide if the pattern is exist or not. The first method we talk about was proposed by Pande et al. [1]. The first step of this method is applying the proposed algorithm to composite the model graph to produce a number of decomposed graphs and then try to applying isomorphism mechanism to find the matches between these decomposed graphs and the design patterns. The drawback of this method is not covering all the 23 patterns of GoF. It is only examined the patterns that match to the composed graph of order 2 or 3 and the complexity of this decomposition algorithm is depend on the number of nodes (it is O (n3) if the graph has three nodes). So, complexity will increased in the systems with large model graph.

Some approaches also decide that if the design pattern partially exists in the system design. To detect the design patterns, Gupta et al. [2] used the Normalized cross correlation (NCC). They started by writing the relationship matrices of the relationship graphs that extracted from the UML diagram of the system design and the design pattern. Then, they apply the NCC to discover the degree of similarity between the two graphs (one for the system design and the other for the design pattern) for the purpose of detecting the Design patterns. The disadvantage of this method is, it is time consuming to calculate each relationship matrix distinctly. Gupta et al [3] proposed a graph matching algorithm based on the A* algorithm. The proposed algorithm divided the matching process, 1) between the design pattern graph (DPG) and the model graph (MG), 2) between the design pattern graph (DPG) and the system graph (SG), into K phases. The value of K is depending on the minimum number of nodes (M) in the two graphs used in matching. It is selected from the values between 1 to M. The Problem with this method, it is effective only in the case of small amounts of K values. Pande et al[4] proposed a new matching algorithm which extracts the relationship directed graphs from the UML diagrams of system model and design patterns. Then they attempt to discover the existent of design patterns in the model graph using Depth-Node-Input Table (DNIT). DNIT arranges the various model graph relationships and the design pattern by depth. This method is time consuming in the process of creating the DNIT entities for the model graph and for all design patterns graphs.

Another approach used the concept of the Boolean function to detect the patterns. Gupta et al. [5] Offering a new method that detect the design pattern depend on the Boolean function. They transform the UML diagrams of the model graph and the design pattern into Boolean function with SOP form. Then, they detect the existence of the design patterns by comparing the two Boolean functions. The drawbacks of using this method are that it not considers the detecting of design patterns that have relationship from a node to itself such as the singleton design pattern.

Some approaches identify the number of design pattern instances in the system. Pande et al. [7] attempted to detect the design patterns and its instances by firstly, construct the decision tree using row-column elements for all possible subgraphs adjacency matrices of the system design graph. After that, find the row-column elements of the design pattern adjacency matrix. Then, check the isomorphic by traverse the decision tree to detect the row-column elements of that design pattern. The limitation of this method appears in the case of constructing a decision tree for large system design graph; it used a huge number of permutations subgraphs and will produce a complex tree that consuming a lot of time in the traversal process. M. Gupta et al. [8], they detect the design pattern by applying a graph matching algorithm where the matching process describes the state space illustration. The graph matching algorithm identifies the complete sub isomorphism for each relationship between two graphs distinctly. Then, they combine these distinct outputs to detect the existence of the design patterns or its alternatives. The mapping process of this method is time consuming. Gupta and Pande [9] propose a new algorithm to detect the design pattern in the source code depend on the subgraph isomorphism relational view. They attempt to find the complete sub isomorphism between the input graph (system under study) and the design pattern graph to detect the existence of the design patterns or its alternatives in the graph of system under study. It is time consuming when find the matches for each relation individually. Yu et al [10] detect the design patterns by graph isomorphism between system design and the design pattern graph. They start by identifying all the candidate nodes from system graph that correspond to the design pattern nodes and then, select some of them to create sub-graphs of system graph. Finally, they find the isomorphic between the design pattern and them. This method discovers how much instance of the design pattern in the system graph. This method represents the relationship between classes in the UML diagram of system model by assigning weight to edges in the graph representation, this weight describe all the relationship between tow vertices. So, in large systems, it will be time consuming to identify the relationship for each edge and then calculate its weight. All the problems of the previous work are summarized in Table 1.

3. **Problem Statement and the proposed Solution**
Following is the problem taken up in this paper based on review of the literature [1-7].

How to detect the existence of the design pattern completely or partially or non-exist in the system design and how many times it exists in the system design?

This proposed solution will illustrate the implementation of the system design and the design pattern UML diagrams into two graphs and the applying the proposed algorithm to detect the completely or partially existence of the design pattern and how many copy of that pattern in the system diagram UML.

3.1 **The implementation of the system design and the design pattern UML diagrams**
First, we implement the system design and the design pattern UML diagrams as two directed graphs. System graph (SG) is corresponded to the system design UML diagram and the design pattern graph (DPG) is corresponded to the design pattern UML diagram. Figures 1 to 5 show the example of system design and design patterns UML with the corresponding SG and DPG respectively.

Each one of these two graphs has set of edges that represent the relationships between tow vertices and the self-loop for a vertex. The SE is the set of edges that implements the SG and the DPE is the set of edges that implements the DPG where each element in the SE and DPE is 4-tuples (A, B, X, Y) .The first 2 elements of the 4-tuples A and B indicate the two vertices of the edge. The third element X indicates the type of relationship between A and B where, value1 indicates the Direct Association, value 2 indicates the dependency and value 3 indicates the generalization relationship. The forth element of the 4-tuples is Y which indicates by 1 if the self-loop existence (if A equal B) otherwise, it will be 0. The advantage of adding the self-loop element is to detect the Singleton Design Pattern where this element indicates if one vertex belongs to itself to make relation. We can simply understand the SE and the DPE sets by implementing these edges by a table of 4 columns to implement A, B, X and Y and N row where N indicates the number of edges in the graph.

3.2 **The New Algorithm to detect the design pattern in the system design**
The proposed algorithm will detect the design pattern depend on find all the possibility of design pattern existence or part of it in the system design by comparing the 4-tuples of SE and DPE to satisfy the value of X and Y. the resulted possibilities must form a connected component of the SG. SE and DPE sets are the Input to the algorithm. The output is the statements that indicate if the pattern exists completely or partially with the number of how much copies of it in the system design. The table Poss was used to save all the possibilities of edges from SE in DPE. The number of columns of Poss indicates the number of edges of the DPE where it will be as the |DPE| if the pattern exists completely or less than |DPE| if the pattern exists partially. The numbers of rows of Poss indicate the number of copies that the design pattern exists completely or partially. Figure 6 shows the steps of the new algorithm.

The output of the proposed algorithm is categorized into one of 3 cases as following subsections:

3.3.1 **The Design pattern exists in the SG with the number of times it exists**
This case will occur if the number of columns of Poss equal to the number of edges of the DPE (i.e. it will be equal |DPE|). The number of how much the pattern completely exists in the SG will indicate by the number of rows of Poss.

If we implement the new algorithm on the System design of Figure 1 to detect the Façade design pattern witch presented with Figure 2 then:
SE= {(a,b,1,0),(c,b,1,0), (a,c,1,0),(d,b,3,0),(e,c,3,0),(d,c,2,0)}

DPE= {(P, Q, 1, 0)}

After applying the algorithm on the SE and DPE, the Poss will contain:

(a,b,1,0)
(c,b,1,0)
(c,a,1,0)

Then the output will be: The design pattern completely exists in the System design with 3 times.

Also, if we implement the new algorithm on the System design of Figure 1 to detect the Prototype design pattern witch presented with Figure 4 then:

SE= {(a,b,1,0),(c,b,1,0), (a,c,1,0),(e,c,3,0),(d,c,2,0)}
DPE= {(b,a,1,0),(c,a,3,0)}

After applying the algorithm on the SE and DPE, the Poss will contain:

(a,b,1,0)        (d,b,3,0)
(a,c,1,0)        (e,c,3,0)
(c,b,1,0)        (d,b,3,0)

Then the output will be: The design pattern completely exists in the System design with 3 times. We exclude the 3 rows of Poss:

(a,b,1,0)        (e,c,3,0)
(a,c,1,0)        (d,b,3,0)
(c,b,1,0)        (e,c,3,0)

because the first two rows are not connected , also the three rows not satisfy the equality condition with the elements of DPE which is the DPE 4-tuple elements has the same node in element 2 (a).

### 3.3.2 The Design pattern partially exists in the SG with the number of times it exists.

This case will occur if the number of columns of Poss less than the number of edges of the DPE (i.e. it will be less than |DPE|) but not equal zero. The number of how much the pattern partially exists in the SG will indicate by the number of rows of Poss.

If we implement the new algorithm on the System design of Figure 1 to detect the Composite design pattern witch presented with Figure 5 then:

SE= {(a,b,1,0),(c,b,1,0), (a,c,1,0),(d,b,3,0),(e,c,3,0),(d,c,2,0)}
DPE={(c,a,1,0),(b,a,3,0),(c,a,3,0)}

Firstly, with n=3 edge the Poss will empty because all the possibility not matching the equality condition with the elements of DPE which is the DPE 4-tuple elements has the same node in element 2 (a). Also, the first two possibilities are not connected.

(e,c,3,0)        (d,b,3,0)        (a,b,1,0)
(e,c,3,0)        (d,b,3,0)        (a,c,1,0)
(e,c,3,0)        (d,b,3,0)        (c,b,1,0)

After that, if we decrease n by 1, we will find 7 possibilities:

(e,c,3,0)        (d,b,3,0)
(d,b,3,0)        (a,b,1,0)
(d,b,3,0)        (c,b,1,0)
(d,b,3,0)        (a,c,1,0)
(e,c,3,0)        (a,b,1,0)
(e,c,3,0)        (c,b,1,0)
(e,c,3,0)        (a,c,1,0)

Then Poss will contain the 3 rows:

(d,b,3,0)        (a,b,1,0)
(d,b,3,0)        (c,b,1,0)
(e,c,3,0)        (a,c,1,0)

The output will be: The design pattern partially exists in the System design with 3 times. The rest 4 rows not matching the equality condition with the elements of DPE which is the DPE 4-tuple elements has the same node in element 2 (a).

### 3.3.3 The Design pattern dose not exists in the SG.
This case will occur if Poss is empty (No rows or columns). Then, the pattern dose not exists in SG. If we implement the new algorithm on the System design of Figure 1 to detect the Singleton design pattern witch presented with Figure 3 then:

SE= {(a,b,1,0),(c,b,1,0),(c,a,1,0),(d,b,3,0),(e,c,3,0),(d,c,2,0)}
DPE= {(A, A, 1, 1)}

After applying the algorithm on the SE and DPE, the Poss will be empty and then the output will be: the design pattern does not exist in the system design.

### 4. Validation Of The Proposed Solution
To validate the new Isomorphic algorithm we using an online questionnaire divided into 4 parts for the goals:
1. The efficiency of the proposed solution: This goal validate that the new Isomorphic graph algorithm is efficient in detecting the design pattern.
2. The usability of the proposed solution: This goal validate that the new Isomorphic graph algorithm is easy to use.
3. The understandability of the proposed solution: This goal validate that the new Isomorphic graph algorithm is easy to understand.
4. The reliability of the proposed solution: This goal validate that the new Isomorphic graph algorithm is covering all the cases about the existence of the design pattern in the system design or not .Also, to validate that this new Isomorphic graph algorithm is able to calculate the number of times the design pattern existence completely or partially in the system design.

Each part consists of 5 -five level scaling questions interrelated to each other to prove the validity of one goal. Liker scale is ranging from 1 to 5 as following:
- very low effect indicating 1;
- low effect indicating 2;
- nominal/average effect indicating 3;
- high effect indicating 4;
- very high effect indicating 5.

The data is analyzed statistically to validate the proposed solution.

### 4.1 Findings
This section will present the findings after analyzing the responses to the questions of each goal individually.

**Cumulative Statistical Analysis of Goal 1 (Efficiency)**

It is essential to prove that the new Isomorphic graph algorithm is efficient in detecting the design pattern. In order to prove the efficiency of the proposed solution, we must firstly prove the efficiency of sub-issues strongly related and affected the efficiency of the new solution. These issues are: the type of algorithm which is an Isomorphic graph Algorithm, the UML diagram implementation for both system design and design pattern as direct graph to represent the relational view as pre-step for the new solution, implementation of edges as a set of 4-tuples elements for both the system graph and design pattern graph, adding the self-loop to facilitate detect the Singleton Design Pattern and the relationship indication elements in implementing edges of both the system graph and design pattern graph.

The responses cumulative analysis results for the goal 1 are summarized in Table 2 and Figure 7 which present the analysis results for the five questions of goal 1.

As we shown from figure 7 and Table 2 about the cumulative descriptive analysis of goal 1, it is clearly that 38.67% of the responses find that the new Isomorphic algorithm is high efficient and 36.67% of them find that our proposed solution is very high efficient in detecting the design pattern in the system design. But, 0.67% of them find this solution low efficient. Where, 24% find the efficient of this solution is normal.

**Cumulative Statistical Analysis of Goal 2 (Usability)**

It is essential to prove that the new Isomorphic graph algorithm is usable in detecting the design pattern. In order to prove the usability of the proposed solution, we must firstly prove the usability of sub-issues strongly related and affected the usability of the new solution. These issues are the same with sub-issues of goal1 that mentioned previously.

The responses cumulative analysis results for the goal 2 are summarized in Table 3 and Figure 8 which present the analysis results for the five questions of goal 2.

As we shown from figure 8 and Table 3 about the cumulative descriptive analysis of goal 2, it is clearly that 44.67% of the responses find that the new Isomorphic algorithm is high usable and 36% of them find that our proposed solution is very high usable in detecting the design pattern in the system design. Where, 19.33% of them find this solution normal usable.

**Cumulative Statistical Analysis of Goal 3 (Understandability)**

It is essential to prove that the new Isomorphic graph algorithm is understandable to detecting the design pattern. In order to prove the understandability of the proposed solution, we must firstly prove the understandability of sub-issues strongly related and affected the understandability of the new solution. These issues are the same with sub-issues of both, goal 1 and 2 that mentioned previously.

The responses cumulative analysis results for the goal 3 are summarized in Table 4 and Figure 9 which present the analysis results for the five questions of goal 3.

As we shown from figure 9 and Table 4 about the cumulative descriptive analysis of goal 3, it is clearly that 41.33% of the responses find that the new Isomorphic algorithm is very high understandable and 40% of them find that our proposed solution is high understandable in detecting the design pattern in the system design. Where, 18.67% of them find this solution normal understandable.

**Cumulative Statistical Analysis of Goal 4 (Reliability)**

It is essential to prove that the new Isomorphic graph algorithm is reliable in detecting the design pattern. In order to prove the reliability of the proposed solution, we must firstly prove the reliability of sub-issues strongly related and affected the reliability of the new solution. These issues are differing from sub-issues of goal 1, 2 and 3 that mentioned previously. These sub issues are: the ability of the proposed solution to detect the completely and partially existence cases, the non-existence of the design pattern in the system design, and its ability to calculate the number of times the design pattern completely or partially existence in system design.

The responses cumulative analysis results for the goal 4 are summarized in Table 5 and Figure 10 which present the analysis results for the five questions of goal 4.

As we shown from Figure 10 and Table 5 about the cumulative descriptive analysis of goal 4, it is clearly that 46% of the responses find that the new Isomorphic algorithm is very high reliable and 44.67% of them find that our proposed solution is high reliable in detecting the design pattern in the system design. Where, 9.33% of them find this solution reliability is normal.

**Cumulative analysis for all the goals of the new solution**

The responses cumulative analysis results for the 4 goals of the proposed solution are summarized in Table 6 and figure 11.

As we shown from Figure 11 and Table 6 about the cumulative descriptive analysis of all the goals, it is clearly that 42% of the responses rank the new Isomorphic algorithm as high and 40% of them rank it as very high scale in detecting the design pattern in the system design .But, 0.17% of them rank it as low scale while, 9.33% of them rank this solution as normal scale.

**5. Conclusion and Future Work**

The authors of this paper proposed a new solution to detect all the possible cases of exist the design pattern or not in the system design. The graph implementation was used to implement both the system design UML diagram and the design pattern UML diagram. Then, the edges for each one of the both two graphs are implemented as a set of 4-tuple elements. Lastly, the new inexact graph isomorphic algorithm was applied on the two sets to detect the design pattern in the system design. This solution solved the problems of the related work and detects the exist or non-exist of all the 23 patterns of GoF and calculate how many times the design pattern exist partially or completely

In the future, the authors plan to implement the proposed solution programmatically as an open source tool to detect the design pattern.

**References**

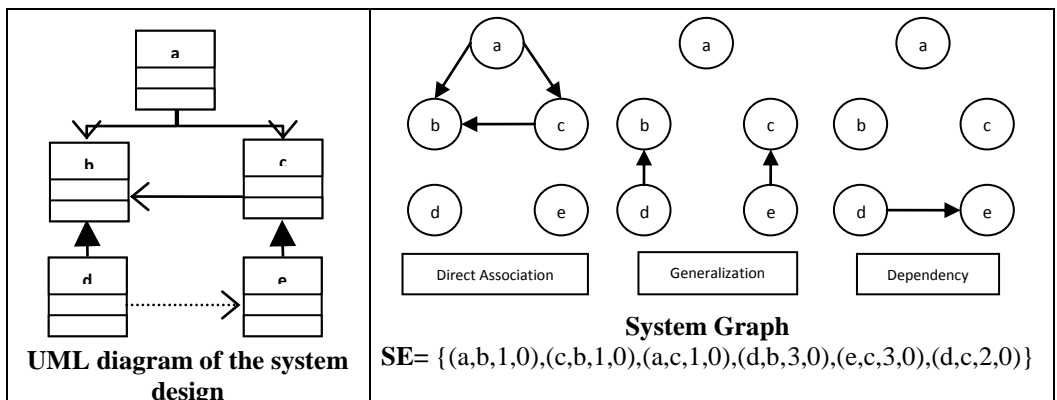

Figure 1 UML diagram of the system design and its SG and SE

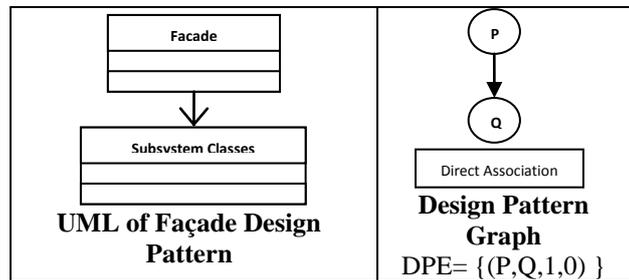

Figure 2 UML diagram of the Façade design pattern and its DPG and DPE

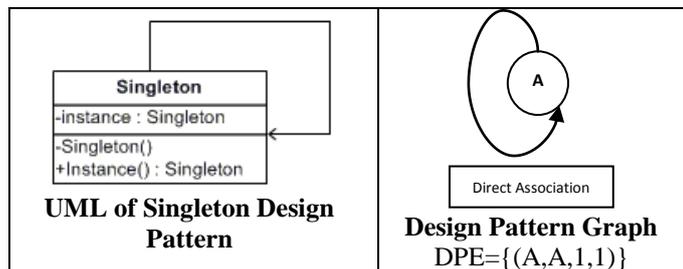

Figure 3 UML diagram of the Singleton design pattern and its DPG and DPE

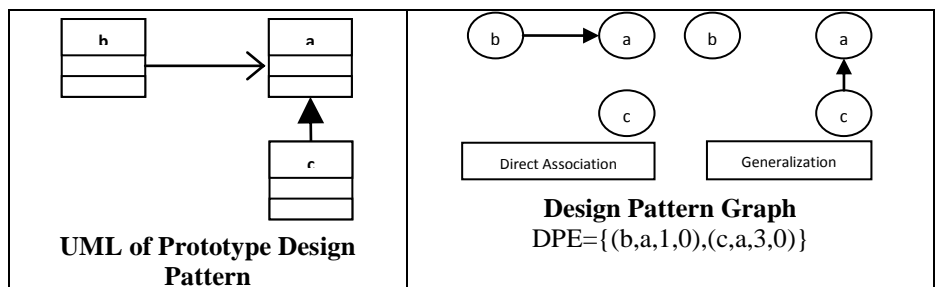

Figure 4 UML diagram of the Prototype design pattern and its DPG and DPE

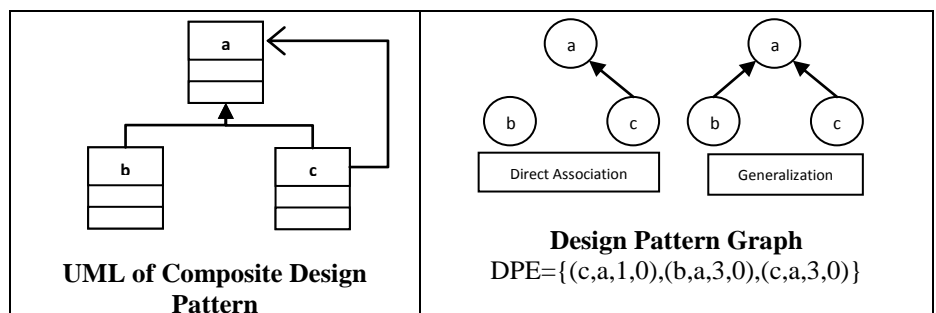

Figure 5 UML diagram of the Composite design pattern and its DPG and DPE

```
Input: SE, DPE
Output: The existence of Design pattern completely or partially with the number of how much copies of it in the SG or not
Steps:
n=|PGE|       //n is the number of DPE elements
|Poss|=0      // the Poss Initial as empty table
    WHILE (n≠0) DO:
    Fined all the possibilities of connected component from SE with n Edges where these edges must satisfy the equality condition with the 4 element of each 4-tuples in DPE
    If |Poss|>0
    Then EXIT WHILE   //the design pattern exist completely or partially
    Else n=n-1    // decrease the number of edge of DPE by 1 to find partially exists of Design Pattern
    END WHILE
Poss_ columns _Count= number of columns of Poss
Poss_ rows _Count= number of rows of Poss
//output
If (Poss_ columns _Count =|DPE|)       //case1
Then: (Print "The design pattern completely exists in the System design with"+ Poss_ rows _Count+" times")
Else If (Poss_ columns _Count <|DPE|) //case2
Then: (Print "The design pattern partially exists in the System design with"+ Poss_ rows _Count+" times")
Else If (Poss_ columns _Count =0 && Poss_ rows _Count=0) //case3
Then: (Print "The design pattern does not exist in the System design")
```

Figure 6 the new algorithm to detect the design pattern in the system design.

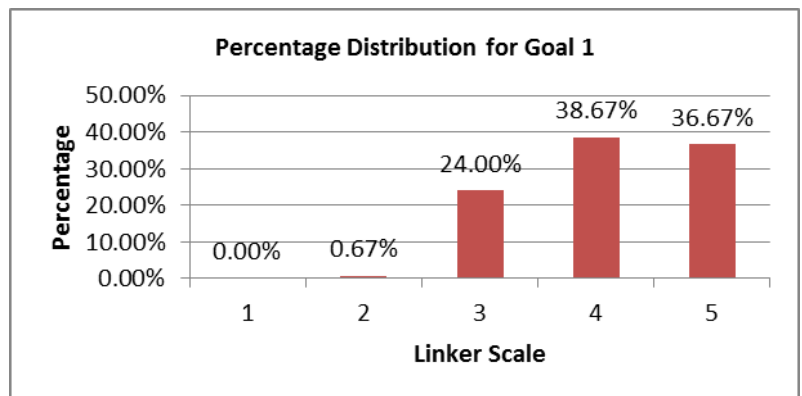

Figure 7 Percentage Distributions for Goal 1

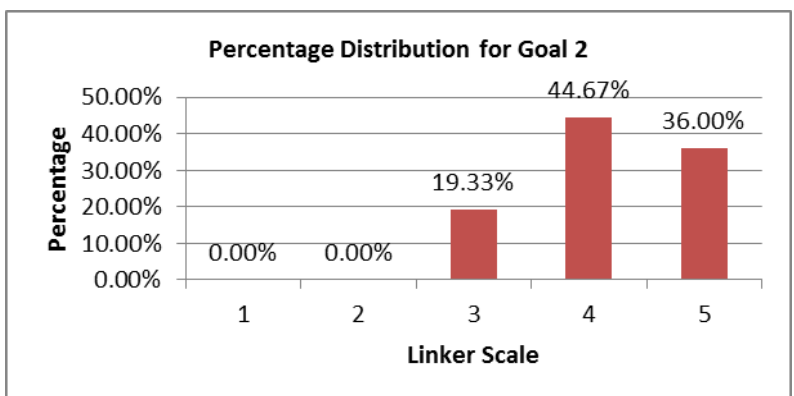

Figure 8 Percentage Distribution for Goal 2

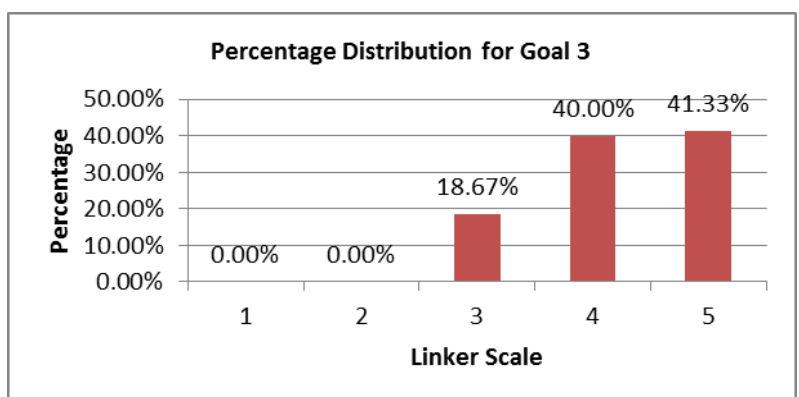

Figure 9 Percentage Distribution for Goal 3

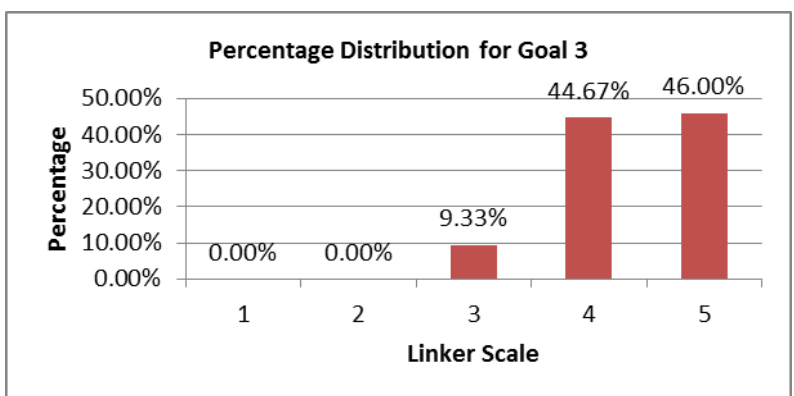

Figure 10 Percentage Distribution for Goal 4

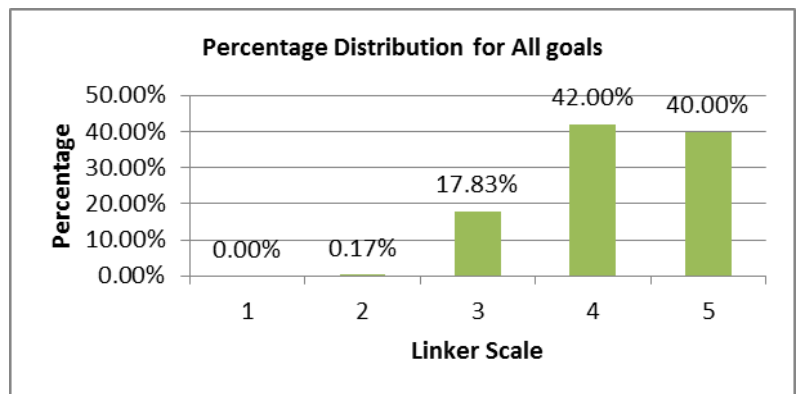

Figure 11 Percentage Distribution for all Goals

Table 1 Summarized related work problems

| Paper Title | Problems |
|---|---|
| A New Approach for Detecting Design Patterns by Graph Decomposition and Graph Isomorphism [1] | • Not cover all the 23 patterns of GoF<br>• Complex for graph with large number of nodes. |
| Design Pattern Detection by Normalized Cross Correlation [2] | • time consuming to calculate each relationship matrix distinctly |
| Design Pattern Detection Using Inexact Graph Matching [3] | • Not efficient when using large values of K. |
| DNIT – A New Approach for Design Pattern Detection [4] | • Time consuming in constructing DNIT |
| Design Pattern Mining for GIS Application Using Graph Matching Techniques [5] | • Can't use in the situation that the system design doesn't have the pattern. |
| Design patterns detection using SOP expressions for graphs [6] | • not identify patterns with self-relationship |
| A Decision Tree Approach for Design Patterns Detection by Subgraph Isomorphism [7] | • for large system design graphs:<br>• Produce complex tree<br>• Time consuming |
| Design Pattern Mining Using State Space Representation of Graph Matching [8] | • The mapping process is time consuming |
| Design Pattern Mining Using State Space Representation of Graph Matching [8] | • Time consuming when find the matches for each relation individually. |
| Design Patterns Mining using Subgraph Isomorphism: Relational View [9] | • Time consuming to identify the relationship for each edge and then calculate its weight. |
| Detection of Design Pattern Instances Based on Graph Isomorphism [10] | • Not cover all the 23 patterns of GoF<br>• Complex for graph with large number of nodes. |

Table 2 the cumulative analysis results for goal 1.

| Q# | Very low | Low | Nominal/Average | High | Very high | Total |
|---|---|---|---|---|---|---|
| Q1 | 0.00% | 3.33% | 23.33% | 36.67% | 36.67% | 100.00% |
| Q2 | 0.00% | 0.00% | 16.67% | 60.00% | 23.33% | 100.00% |
| Q3 | 0.00% | 0.00% | 20.00% | 33.33% | 46.67% | 100.00% |

| Q4 | 0.00% | 0.00% | 30.00% | 30.00% | 40.00% | 100.00% |
| Q5 | 0.00% | 0.00% | 30.00% | 33.33% | 36.67% | 100.00% |
| Total | 0.00 | 3.33 | 120.00 | 193.33 | 183.33 | 500.00 |
| Avg. | 0.00% | 0.67% | 24.00% | 38.67% | 36.67% | 100.00% |

Table 3 the cumulative analysis results for goal 2.

| Q # | Very low | Low | Nominal/Average | High | Very high | Total |
|---|---|---|---|---|---|---|
| Q6 | 0.00% | 0.00% | 20.00% | 53.33% | 26.67% | 100.00% |
| Q7 | 0.00% | 0.00% | 16.67% | 43.33% | 40.00% | 100.00% |
| Q8 | 0.00% | 0.00% | 16.67% | 50.00% | 33.33% | 100.00% |
| Q9 | 0.00% | 0.00% | 23.33% | 23.33% | 53.33% | 100.00% |
| Q10 | 0.00% | 0.00% | 20.00% | 53.33% | 26.67% | 100.00% |
| Total | 0.00 | 0.00 | 96.67 | 223.33 | 180.00 | 500.00 |
| Avg. | 0.00% | 0.00% | 19.33% | 44.67% | 36.00% | 100.00% |

Table 4 the cumulative analysis results for goal 3.

| Q # | Very low | Low | Nominal/Average | High | Very high | Total |
|---|---|---|---|---|---|---|
| Q11 | 0.00% | 0.00% | 16.67% | 43.33% | 40.00% | 100.00% |
| Q12 | 0.00% | 0.00% | 13.33% | 53.33% | 33.33% | 100.00% |
| Q13 | 0.00% | 0.00% | 16.67% | 40.00% | 43.33% | 100.00% |
| Q14 | 0.00% | 0.00% | 20.00% | 26.67% | 53.33% | 100.00% |
| Q15 | 0.00% | 0.00% | 26.67% | 36.67% | 36.67% | 100.00% |
| Total | 0.00 | 0.00 | 93.33 | 200.00 | 206.67 | 500.00 |
| Avg. | 0.00% | 0.00% | 18.67% | 40.00% | 41.33% | 100.00% |

Table 5 the cumulative analysis results for goal 4.

| Q # | Very low | Low | Nominal/Average | High | Very high | Total |
|---|---|---|---|---|---|---|
| Q16 | 0.00% | 0.00% | 16.67% | 40.00% | 43.33% | 100.00% |
| Q17 | 0.00% | 0.00% | 16.67% | 30.00% | 53.33% | 100.00% |
| Q18 | 0.00% | 0.00% | 6.67% | 50.00% | 43.33% | 100.00% |
| Q19 | 0.00% | 0.00% | 3.33% | 46.67% | 50.00% | 100.00% |
| Q20 | 0.00% | 0.00% | 3.33% | 56.67% | 40.00% | 100.00% |
| Total | 0.00 | 0.00 | 46.67 | 223.33 | 230.00 | 500.00 |
| Avg. | 0.00% | 0.00% | 9.33% | 44.67% | 46.00% | 100.00% |

Table 6 the cumulative analysis results for all goals.

| Q # | Very low | Low | Nominal/Average | High | Very high | Total |
|---|---|---|---|---|---|---|
| Goal 1 | 0.00% | 3.33% | 120.00% | 193.33% | 183.33% | 500.00% |
| Goal 2 | 0.00% | 0.00% | 96.67% | 223.33% | 180.00% | 500.00% |
| Goal 3 | 0.00% | 0.00% | 93.33% | 200.00% | 206.67% | 500.00% |
| Goal 4 | 0.00% | 0.00% | 46.67% | 223.33% | 230.00% | 500.00% |
| Total | 0.00 | 3.33 | 356.67 | 840.00 | 800.00 | 2000.00 |
| Avg. | 0.00% | 0.17% | 17.83% | 42.00% | 40.00% | 100.00% |

**Authors' profiles**

**M. Rizwan Jameel Qureshi-** Assistant Professor at Faculty of Computing & Information Technology, King Abdulaziz University, major in CBD and agile

**Afnan Salem Babrahem-** Master student in IT Department at King Abdulaziz University, interested in Technology Management, Computer Networks and Global Software Engineering.